\documentclass[fleqn,twoside]{article}
\usepackage{espcrc2}

\usepackage{graphicx}
\usepackage{epsfig}
\usepackage[figuresright]{rotating}

\newcommand{\AmS}{{\protect\the\textfont2
  A\kern-.1667em\lower.5ex\hbox{M}\kern-.125emS}}

\def\dmsq23{\Delta m^2_{32}}

\def\numu{\nu_\mu}

\def\nue{\nu_e}
\def\sinsq2t23{\sin^2 2\theta_{23}}
\def\ssq2t13{\sin^2 2\theta_{13}}
\def\th13{\theta_{13}}
\def\numutonue{\nu _{\mu }\rightarrow \nu _{e}}

\title{Neutrino detectors for future experiments}

\author{Andr\'e Rubbia\address[ETHZ]{Institute for Particle Physics, ETH-Zurich\\
CH-8093 Zurich, Switzerland}%
        \thanks{This work was supported by ETHZ and Swiss National Science Foundation.}
        }
       
\begin{document}

\begin{abstract}
We review detector technologies which are currently considered for
ultimate nucleon decay searches,
new generation astrophysical neutrinos studies,
and for future long-baseline neutrino
experiments at new high-intensity neutrino beam facilities. We focus our discussion on
Phase-II experiments with a timescale of $10\simeq 20$~years. We point out that there
are very few detector technologies which are general purpose and versatile
enough in order to potentially address non-accelerator based physics 
and a large variety of future neutrino beam types ranging from superbeams,
betabeams and neutrino factories from subGeV to 10's GeV energies.
\vspace{1pc}
\end{abstract}

\maketitle

\section{NEW GENERATION DETECTORS}

It is expected that new neutrino facilities will provide new opportunities
to further develop neutrino physics, especially when coupled to long
baselines. 

The current round of long baseline experiments (``phase-0'') are meant to confirm the 
neutrino oscillation effect found in atmospheric neutrinos.
K2K\cite{Oyama:1998bd}, MINOS\cite{Gallagher:2004pj}, OPERA\cite{Kodama:1998fx} 
and ICARUS\cite{Aprili:2002wx} belong to this generation.

After that, the focus will shift towards second generation experiments (``phase-I'')
designed and optimized to
detect the subleading $\nu_\mu\rightarrow \nu_e$ atmospheric oscillations.
The T2K experiment in Japan is the only approved project, with its beamline
under construction and commissioning planned for 
$\simeq 2009$. The T2K beam is expected to reach
a proton intensity of 0.75~MW around 2011 and could be upgraded
to 4~MW by increasing the repetition rate of the 50~GeV synchrotron at J-PARC,
eliminating idle time in the acceleration cycle and by doubling
the number of circulating protons\cite{kobayashi}. The target region is however
a critical item for the high intensity. The far detector is SuperKamiokande.
T2K should measure $\nu_\mu\rightarrow\nu_\mu$ with $\delta(\Delta m^2_{23})\approx10^{-4} eV^2$ and $\delta(\sin^2\theta_{23})\approx0.01$.
It will find evidence for $\nu_\mu\rightarrow\nu_e$ if $\sin^22\theta_{13}>0.006$.

With Phase 0\&1 experiments on their way, one must consider possible
``phase-II'' projects in $10\simeq 20$ years (?), able to
address more aspects of the phenomenology of neutrino oscillations, beyond
those addressed by the currently approved experiments. For example:
 
\begin{table*}[htb]
\caption{Quick comparison of detector technologies for future experiments. Solar means
solar neutrinos. SN means supernovae neutrinos (burst + relic). Atm means atmospheric
neutrinos.}
\label{table:futdet}
\newcommand{\m}{\hphantom{$-$}}
\newcommand{\cc}[1]{\multicolumn{1}{c}{#1}}
\begin{tabular}{@{}|l|c|c|c|c|c|c|c|c|c|}
\hline
Detector & Mass & Solar & SN & Atm & Nucleon & \multicolumn{3}{|c|}{Superbeam, $\beta$-beam} & $\nu$-factory \\
 & kt & & & & decay  &  subGeV & GeV & 10's GeV & 10's GeV \\
\hline
WC & $\simeq 1000$ & $\approx$ & {\bf yes} & {\bf yes} & {\bf yes} & {\bf yes} & $\approx$ & no & no \\
\hline
LAr & $\simeq 100$ & {\bf yes} & {\bf yes} & {\bf yes} & {\bf yes} & {\bf yes} & {\bf yes} & {\bf yes} &  {\bf yes} ($\mu$-catcher) \\
\hline
Magnetized LAr & $\simeq 25$ & {\bf yes} & {\bf yes} & {\bf yes} & {\bf yes} & {\bf yes} & {\bf yes} & {\bf yes} &  $e^\pm$, $\mu^\pm$ \\
\hline
Magnetized & $\simeq 50$ & no & no & $\mu^\pm$ & no & $\approx$ & {\bf yes} & {\bf yes} &  $\mu^\pm$ \\
sampling Cal. & & & & & & & & & \\
\hline
Non-magnetized & $\simeq 50$ & no & no & $\mu$'s & no & $\approx$ & {\bf yes} & {\bf yes} &  no \\
sampling Cal.  & & & & & & & & & \\
\hline
Emulsion & $\simeq 1$ & no & no &no & no & no & $\approx$ & {\bf yes} &  $\tau^\pm$ \\
hybrid   & & & & & & & & & \\
\hline
\end{tabular}\\[2pt]
\end{table*}

\begin{enumerate}
\item Study precisely the $L/E$ dependence of the oscillation probability;
\item Improve errors to $\delta(\Delta m^2_{23})\approx 1\%$;
\item Improve errors to $\delta(\sin^2\theta_{23})<1\%$;
\item Improve sensitivity to $\sin^22\theta_{13}$  by a factor $\times 5$ or $\times 10$ w.r.t. to T2K or precisely measure it 
if T2K has found a signal;
\item Find evidence for CPV ($\delta\ne 0$);
\item Fix the sign of $\Delta m^2_{23}$  and study matter effects;
\item Observe ($\Delta m^2_{21}, \sin^2\theta_{12}$) oscillations in terrestrial experiments;
\item Over-constrain the $U_{PMNS}$ matrix (unitarity tests);
\item Search for non-standard interactions;
\item Any other business.
\end{enumerate}

For the search for CP violation in the lepton sector
and/or the determination of the mass hierarchy, the discovery of  a non-vanishing
$\theta_{13}$ via  the subleading $\nu_\mu\rightarrow \nu_e$ oscillation in the
Phase-I is a prerequisite. Otherwise, the
Phase-II is designed exclusively to improve the sensitivity
to $\nu_\mu\rightarrow \nu_e$ oscillations with little other prospects. Note that for $\sin^22\theta_{13}<\simeq 0.001$,
a non-negligible amount of $\nu_\mu\rightarrow \nu_e$ transitions are induced by the
solar parameters ($\Delta m^2_{12}$  , $\theta_{12}$) hence energy dependent studies of the
oscillation probability will be mandatory to disentangle the $1-2$ and $1-3$-driven
effects. 

Since not all parameters relevant for the physics of Phase-II are known (e.g. $\theta_{13}$),
we should tend towards a general purpose and versatile
detector that gives us the largest opportunity to perform interesting and new physics 
in the year $\simeq 2020$, whatever that physics will be.
It appears that such a detector should possess the following attributes:
\begin{enumerate}
\item  Should be very massive \& general purpose, and not solely
``tuned'' to a given physics topic which might be relevant today, but not necessarily tomorrow;
\item  Should detect a wide range of energies and 
have the proper energy resolution to ``see the oscillations'', measure the oscillations parameters precisely and disentangle possible degeneracy;
\item  Should have the granularity to potentially address all the existing $e/\mu/\tau$ flavors in the final states;
\item  Should have a clean NC, CC separation and good background suppression;
\item  Should address both accelerator \& non-accelerator physics, hence be  located underground (depth to be optimized);
\item  Should be ready to find the unexpected (many years will pass from design to data takingÉ);
\item  Should be cost effective.
\end{enumerate}
It is also clear that one needs to consider the complete system including
detector, accelerator complex and  beam type simultaneously,
and systematically optimize the energies, baselines, intensities, ...
This challenging task has not been yet fully completed given the size of
the parameter space. It is important to pursue such optimizations (``feasibility studies'')
before embarking on a specific road for Phase-II projects.

We believe that new experiments of the envisaged scale must address a wide non-accelerator physics program as well as, either 
independently while waiting to be coupled to a Phase-II neutrino facility, or
simultaneously performing accelerator and non-accelerator physics programs. Hence, it is important to develop new 
massive underground detectors for astrophysical neutrinos and nucleon decay searches, 
while keeping the possibility open of a future neutrino facility directed towards it. 
In general, laboratories planning new accelerator developments and/or upgrades
should keep in mind possible neutrino physics.

A comparison of possible detector technologies is summarized in Table~\ref{table:futdet}, organized
according to their types and desired masses. We consider Water Cerenkov detectors (WC) at the megaton-scale,
liquid Argon TPC's without and with magnetic field (LAr), fine sampling calorimeters with and 
without magnetic field, and emulsion hybrid detectors.
The 3th to 6th columns illustrate the potentialities
to address the non-accelerator physics program (solar, supernova, atmospheric neutrinos and nucleon
decay searches). The last columns illustrate the capability to study artificial neutrino beams
like Superbeam or Betabeams of various energies (subGeV, GeV and 10's GeV) and the neutrino
factory (assumed to have 10's GeV energy).

The various detector technologies are discussed in more details in the following sections.

\section{WATER CERENKOV DETECTORS}
Two generations of large water Cerenkov detectors at Kamioka (Kamiokande\cite{Koshiba:mw} and Superkamiokande\cite{Fukuda:2002uc})
have been very successful in research of neutrino physics with astrophysical sources. In addition,
the first long baseline neutrino oscillation experiment with accelerator-produced neutrinos, K2K,
has been conducted with Superkamiokande as far detector. 

Superkamiokande is composed of a tank of 50~kton of water (22.5~kton fiducial) which is
surrounded by 11146 20-inch phototubes immersed in the water. About $170~\gamma/cm$ are produced
by relativistic particles in water in the visible wavelength $350<\lambda<500~nm$. With 40\% PMT coverage
and a quantum efficiency of 20\%, this yields $\approx 14$ photoelectrons per cm or $\approx 7$~p.e. per MeV 
deposited.

\begin{figure}[tb]
\centerline{\epsfxsize=0.5\textwidth\epsfbox{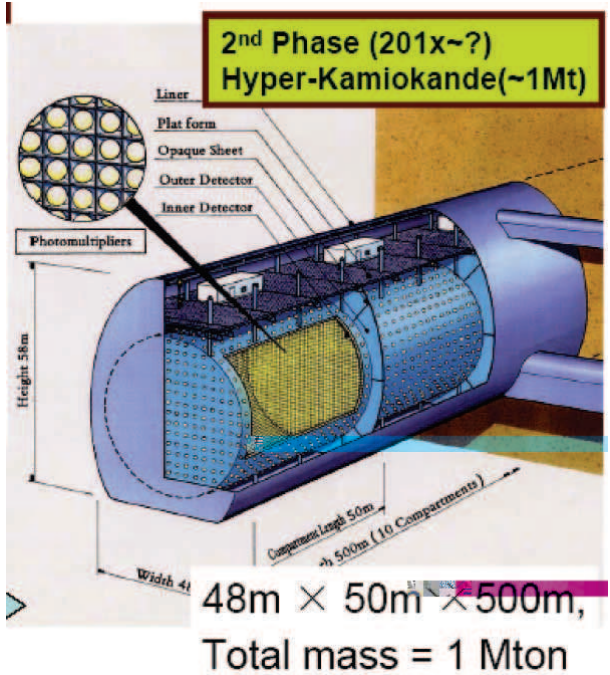}}   
\caption{Possible configuration of the Hyper-Kamiokande detector.}
\label{fig:hyperk}
\end{figure}

There are good reasons to consider a third generation water Cerenkov detector with an
order of magnitude larger mass than Superkamiokande: Hyperkamiokande~\cite{Nakamura:2003hk} (See Figure~\ref{fig:hyperk}) has been
proposed with about 1~Mton, or about 20 times as large as Superkamiokande, based
on a trade-off between physics reach and construction cost. 
Further scaling is limited by light propagation in water (scattering, absorption).

A megaton Water Cerenkov detector will have a 
broad physics programme, including both
non-accelerator (proton decay, supernovae, É) and accelerator physics.

The concept of Hyper-Kamiokande
was initially developed for nucleon decay searches. However, the possibility to
use it as a far detector of a new high-intensity neutrino facility was considered to
be an important purpose. Recently, the approval of the T2K project with 
a high intensity neutrino beam towards Superkamiokande
has reopened the possibility to use HyperKamiokande as far detector to measure CP-violation
in the leptonic sector. Indeed, with the 4~MW upgrade of the synchrotron and the 25 times
larger fiducial mass, statistics at HK will be 100 times higher than at T2K-SK Phase-I.

Although this order of magnitude
extrapolation in mass is often considered as straight-forward, a number of R\&D efforts including the site
selection are needed before designing the real detector. At present, the following items
are being studied: (i) site selection, (ii) cavity design and assessment, (iii) detector tank
design and study of construction method, (iv) simulation studies of the detector performance
for nucleon decay search and long baseline neutrino oscillation experiment, and (iv) development
of HPDs (hybrid photo-detectors).

The Mozumi Mine, which is the current Superkamiokande site, cannot accomodate HyperK
because there is no region of stable hard rock wide enough. Even if such a region existed, large-scale
blasting for the excavation of a very big cavity should be avoided around Super-K which must
stay in operation. A new site in Tochibora at a depth of 1400-1900 m.w.e. was found, which is 
located about 8~km south of the Mozumi Mine. This location allows for a solution to provide
the T2K neutrino beam with the same spectral properties to both Super-K and Hyper-K.

\begin{figure}[tb]
\centerline{\epsfxsize=0.5\textwidth\epsfbox{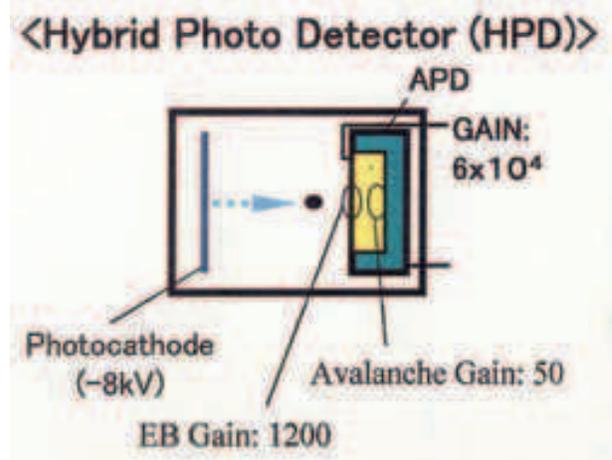}}   
\caption{Schematic of a prototype 5-inch HPD.}
\label{fig:hpd}
\end{figure}

The inner detector of Superkamiokande is instrumented with Hamamatsu 20-inch PMTs.
An important item for Hyperkamiokande is developments of new photo-detectors: with the
same photo-sensitive coverage as that of Superkamiokande ($\approx 2 PMTs/m^2$), 
the total number of PMTs needed for HyperK will be $\simeq 200'000$. Possibilities
to have devices with higher quantum efficiency, flat or thin and possibly operating in
magnetic field are being pursued. Before the Superkamiokande accident, development
of PMTs larger than 20 inch was seriously considered. However, R\&D in this direction
was reduced due to reasons of safety. Current efforts are focused on developments
of large HPDs in collaboration with Hamamatsu. HPDs have a simpler structure than
PMTs, which may allow for a pressure-tight spherical spherical shape. Also, their
production would be easier and cheaper than PMTs of similar size. Current goals
are to reach 20-inch spherical HPDs. So far, 5-inch HPDs (see Figure~\ref{fig:hpd})
have been prototyped and successfully tested. An avalanche gain of 50 and an electron
bombarded gain of 1000 (3000) were obtained with HV at 8~kV (16~kV). Therefore,
a total gain of $5\times 10^4$ was obtained at 8~kV. The next stage is the development
of 13-inch HPDs.

\begin{figure*}[tb]
\centerline{\epsfxsize=0.95\textwidth\epsfbox{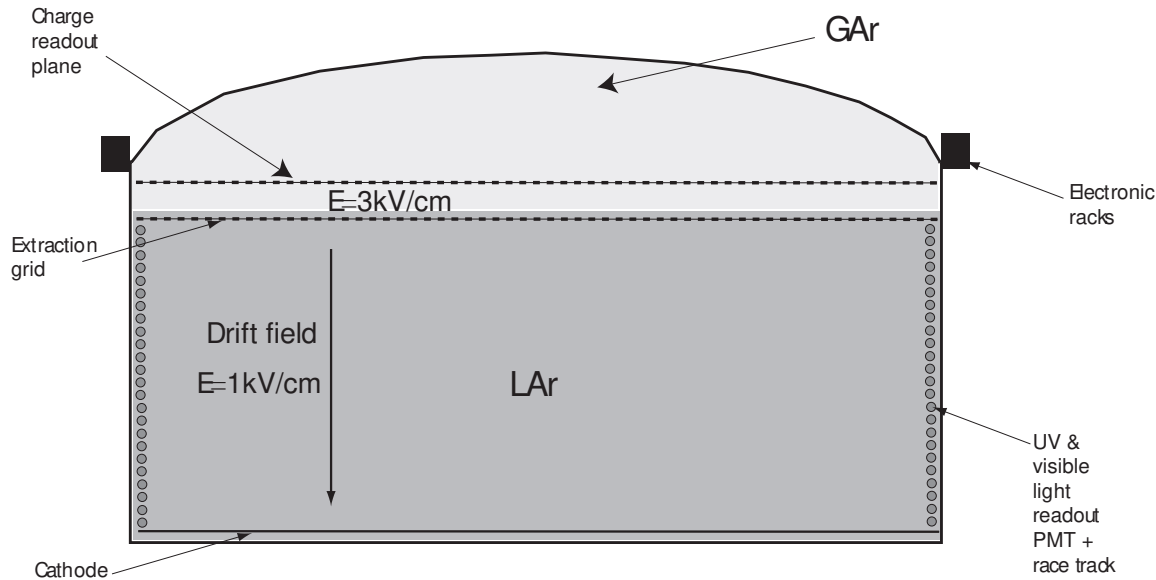}}   
\caption{Schematic layout of a 100 kton liquid Argon detector. The race track is composed of a set
of field shaping electrodes.}
\label{fig:t100schema}
\end{figure*}

Another option is the UNO detector\cite{Jung:1999jq}
with the currently favored location at the Henderson mine (USA).

Finally a megaton detector has been proposed for the Fr\'ejus tunnel in Europe.

\section{LIQUID ARGON TIME PROJECTION CHAMBER}
\subsection{The technique}
Among the many ideas developed around the use of liquid noble gases, the Liquid 
Argon Time Projection Chamber~\cite{intro1} certainly represented one of the most
challenging and appealing designs.
The technology was proposed as a tool for
uniform and high accuracy imaging of massive detector volumes. 
The operating principle of the LAr TPC was based on
the fact that in highly purified LAr ionization tracks can be transported
undistorted by a uniform electric field over distances of the 
order of meters~\cite{Aprile:1985xz}. Imaging is
provided by wire planes placed at the end of
the drift path, continuously sensing and recording the signals induced by
the drifting electrons. Liquid Argon is an ideal medium since it provides
high density, excellent properties (ionization, scintillation yields) and
is intrinsically safe and cheap, and readily available anywhere as a standard by-product
of the liquefaction of air.

The feasibility of this technology has been
demonstrated by the extensive ICARUS R\&D programme, which included
studies on small LAr volumes about proof of principle, LAr purification
methods, readout schemes and electronics, as well as studies with
several prototypes of increasing mass on purification technology,
collection of physics events, pattern recognition, long duration tests and
readout. The largest of these devices had a mass of 3 tons of
LAr~\cite{3tons,Cennini:ha} and has been continuously operated for more than four years, collecting
a large sample of cosmic-ray and gamma-source events. Furthermore, a smaller
device with 50 l of LAr~\cite{50lt} was exposed to the CERN neutrino
beam, demonstrating the high recognition capability of the technique for
neutrino interaction events.

The largest liquid Argon TPC ever build so far is the ICARUS T600 detector, whose
successful assembly culminated with its full test 
carried out at surface during the summer 2001~\cite{Amerio:2004ze,Amoruso:2003sw,Amoruso:2004dy,gg2,gg3,gg1}. 
Installation of this detector at the LNGS is ongoing and commissioning for data taking
is expected in 2006.

A 100 kton liquid Argon TPC would deliver extraordinary physics output, owing to better event reconstruction
capabilities provided by the LAr technique and would be 
one of the most advanced massive underground detectors
built so far~\cite{Rubbia:2004yq} with a rich astrophysical
and accelerator based physics program: detection of supernova neutrinos\cite{Gil-Botella:2004bv},
signal from relic supernova\cite{Cocco:2004ac}, oscillation
physics at future neutrino beam facilities\cite{Rubbia:2001pk,ECT04,Bueno:2001jd,Bueno:2000fg,Rubbia:2004tz}.

In order to reach the 10-100~kton mass adequate for a Phase-II, a new concept
is required to extrapolate further the technology. Such a conceptual design
is outline in the next section.

\subsection{A conceptual design for a 100~kton detector}
A conceptual design for a 100~kton LAr TPC was first given in Ref.~\cite{Rubbia:2004tz} (See Figure~\ref{fig:t100schema}).
The basic design features of the detector can be summarized as follows:
(1) Single 100 kton ``boiling'' cryogenic tanker at atmospheric
pressure for a stable and safe equilibrium condition (temperature is constant while Argon is boiling).
The evaporation rate is small (less than $10^{-3}$ of the total volume per day given
by the very favorable area to volume ratio) and is compensated
by corresponding refilling of the evaporated Argon volume. 
(2) { Charge imaging, scintillation and Cerenkov light readout}
for a complete (redundant) event reconstruction\cite{Rubbia:2004tz} . This represents a clear advantage over large
mass, alternative detectors operating with only one of these readout modes. 
Scintillation and Cerenkov light can be readout essentially independently for improved
physics performance\cite{ECT04}.
(3)  { Charge amplification to allow for very long drift paths}. 
The detector is running in bi-phase mode. In order to allow for drift lengths as long as $\sim$ 20 m,
which provides an economical way to increase the volume of the detector with a constant number
of channels, charge attenuation will occur along the drift due to attachment to the remnant impurities present
in the LAr. This effect can be compensated with 
charge amplification near the anodes located in the gas phase.

The cryogenic features of the proposed design are based on the industrial
know-how in the storage of liquefied natural gases (LNG, $T\simeq 110$ K at 1 bar),
which developed quite dramatically in the last decades, driven by the petrochemical and space rocket industries. 
LNG are used when volume is an issue, in particular, for storage.
The technical problems associated to the design of large cryogenic tankers,
their construction and safe operation have already been addressed and engineering problems
have been solved by the petrochemical industry. 
The current state-of-the-art contemplates cryogenic tankers of
200000~m$^3$ and their number in the world is estimated to be $\sim$~2000
with volumes larger than 30000~m$^3$ with the vast majority built
during the last 40 years. 
Technodyne International Limited, UK~\cite{Technodyne}, which
has expertise in the design of LNG tankers, has been appointed to initiate a feasibility
study in order to understand and clarify the issues related to the operation of a large
underground LAr detector. Their final report~\cite{technorep} demonstrates
the technical feasibility and cost of this LAr tanker for physics experimentation.

A schematic layout of the inner detector is shown in Figure~\ref{fig:t100schema}. The detector is characterized
by the large fiducial volume of LAr included in a large tanker, with external dimensions
of approximately 40 m in height and 70 m in diameter. A cathode located at the bottom of the 
inner tanker volume 
creates a drift electric field of the order of 1~kV/cm over a distance of about 20~m. 
In this field configuration ionization electrons
are moving upwards while ions are going downward. The electric field is delimited on the sides of the tanker
by a series of ring electrodes (race-tracks) placed at the appropriate potential by a voltage divider.

The tanker contains both liquid and gas Argon phases at equilibrium. Since purity is a concern for very long
drifts of 20 m, we assume that the inner detector could be operated in bi-phase mode:
drift electrons produced in the liquid phase are extracted from the liquid into the gas phase with
the help of a suitable electric field and then amplified near the anodes. 
In order to amplify the extracted charge one can consider various options: amplification
near thin readout wires, GEM~\cite{Sauli:qp} or LEM~\cite{Jeanneret:mr}. 
Studies that we are presently conducting show that gain factors of 100-1000 are achievable in pure Argon~\cite{dmrd}.
Amplification operates in proportional mode. Since the readout is limited to the top of the detector, 
it is practical to route cables out from
the top of the dewar where electronics crates can be located around the dewar outer edges.

\begin{figure}[tb]
\begin{center}
\epsfig{file=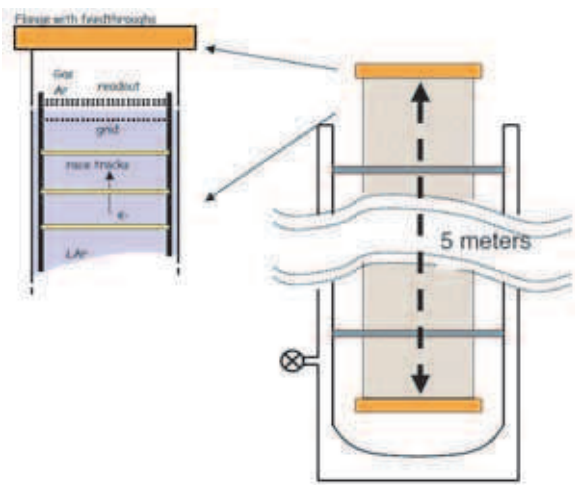,height=0.4\textwidth} 
 \caption{A 5 m long detector column is being realized as part of the R\&D activity
for very long drift paths with charge extraction and amplification.}
\label{fig:pistolone}
\end{center}
\end{figure}

After a drift of 20 m at 1 kV/cm, the electron cloud
diffusion reaches approximately a size of 3 mm, which corresponds to the
envisaged readout pitch. Therefore, 20 m practically
corresponds to the longest conceivable drift path. 
As mentioned above, drifting over such distances
will be possible allowing for some charge attenuation due to attachment
to impurities. If one assumes that the operating electron lifetime is at least $\tau\simeq 2$~ms,
one then expects an attenuation of a factor $\sim$ 150 over the distance of 20~m. 
This loss will be compensated by the proportional gain at the anodes.
We remind that the expected attenuation factor (compensated by the amplification) will not introduce
any detection inefficiency, given the value of $\sim$ 6000 ionization electrons per millimeter produced 
along a minimum ionizing track in LAr.

In addition to charge readout, one can envision to locate PMTs around the inner surface of the tanker. 
Scintillation and Cerenkov light can be readout essentially independently. 
LAr is a very good scintillator with about 50000 $\gamma$/MeV (at zero electric field). 
However, this light is essentially distributed around a line at $\lambda=128$~nm and, therefore, 
a PMT wavelength shifter (WLS) coating is required. 
Cerenkov light from penetrating muon tracks has been successfully detected in a LAr TPC~\cite{gg2};
this much weaker radiation (about $700\ \gamma/$MeV between 
160~nm and 600~nm for an ultrarelativistic muon) can be separately identified with
PMTs without WLS coating, since their efficiency for the DUV light
will be very small. 

A series of R\&D is ongoing to further develop the conceptual ideas outlined above,
with the aim of identifiying the main issues 
of the future systematic R\&D and optimization activities\cite{Ereditato:2004ru}: 
(1) The study of suitable charge extraction, amplification and collection devices;
(2) The understanding of charge collection under high pressure as expected for events occurring at the bottom of the
large cryogenic tanker;
(3) The realization of a 5 m long detector column\cite{Ereditato:2004ru} (See Figure~\ref{fig:pistolone}):
We are constructing a column-like dewar 6 m long and 40 cm in diameter which will contain a 5 m long prototype
LAr detector. The device will be operated with a reduced electric field value
in order to simulate very long drift distances of up to about 20 m.
Charge attenuation and amplification will be studied in detail together with the adoption of possible novel technological
solutions. In particular,
several options are being studied for both the HV field shaping electrodes and for the readout devices.

\subsection{A magnetized liquid Argon TPC}
The possibility to complement the LAr TPC with those
provided by a magnetic field has been considered and would
open new possibilities, important in the case of a neutrino factory\cite{Rubbia:2001pk,Rubbia:2004tz}:
(a) charge discrimination,
(b) momentum measurement of particles escaping the detector ($e.g.$ high energy muons),
(c) very precise kinematics, since the measurements are multiple scattering
dominated (e.g. $\Delta p/p\simeq 4\%$ for a track length of $L=12\ m$ and
a field of $B=1T$).

Unlike muons or hadrons, the early showering of electrons 
makes their charge identification difficult. The track length
usable for charge discrimination is limited to a few radiation
lengths after which the showers makes the recognition of
the parent electron more difficult. In practice, charge discrimination
is possible for high fields $x=1X_0 \rightarrow B>0.5T$, $x=2X_0 \rightarrow B>0.4T$,
$x=3X_0 \rightarrow B>0.3T$.
From simulations, we found that the determination
of the charge of electrons of energy in the range between
1 and 5 GeV is feasible with good
purity, provided the field has a strength in the range of 1~T.
Preliminary estimates show that
these electrons exhibit an average curvature 
sufficient to have electron charge discrimination better than
$1\%$ with an efficiency of 20\%\cite{Bueno:2001jd}. Further studies are on-going.

\begin{figure}[tb]
\begin{center}
\epsfig{file=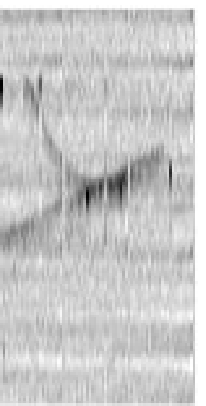,height=0.4\textwidth} 
\epsfig{file=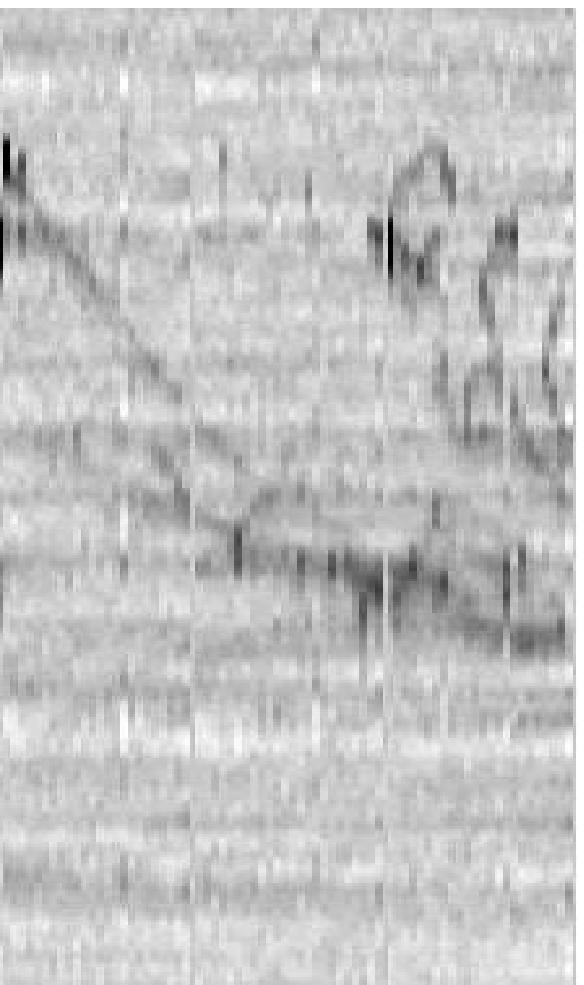,height=0.4\textwidth} 
 \caption{Examples of real events collected\cite{bmaglar} with a small liquid Argon TPC prototype immersed in a magnetic field of 0.55~T:
 (left) bending of delta-ray (right) opening of an $e^+e^-$ pair.}
\label{fig:laff3}
\end{center}
\end{figure}

An R\&D program to investigate a LAr drift chamber in a magnetic
field was started. 
In November 2004 the setup was finalized and first tests could be performed\cite{bmaglar}. Following a cool-down phase of a few
days, the chamber was filled up with liquid Argon and very clean cosmic ray tracks could be observed. After a few
days of commissioning, the magnetic field was turned on and events collected with a liquid Argon TPC immersed
in a magnetic field were collected (see Figure~\ref{fig:laff3}) demonstrating that it is possible
to have a detector with the full bubble-chamber-like fine grain resolution provided by the liquid Argon imaging, with the additional possibility to 
measure particle momenta and the sign of electric charge via magnetic bending. 

In order to fully address the oscillation processes at a neutrino
factory, a detector should
be capable of identifying  and measuring all three charged lepton flavors
produced in charged current interactions {\it and} of measuring 
their charges to discriminate the incoming neutrino helicity. This would
be the case with a magnetized liquid Argon TPC.
From quantitative analyses of neutrino oscillation scenarios\cite{Bueno:2000fg}, one found
that in many cases the discovery sensitivities and the measurements of
the oscillation parameters
were dominated by the ability to measure the muon charge.
However, there were many cases where identification of electron and tau samples 
contributes significantly (see Refs.~\cite{Bueno:2001jd,Bueno:2000fg}). For example, it would open the possibility
to measure the electron charge vital for the search for $T$-violation!

\section{NON-MAGNETIZED FINE-SAMPLING CALORIMETER}
The NuMI neutrino beam line and the MINOS experiment\cite{Gallagher:2004pj} represent a major 
investment of US High Energy Physics 
in the area of neutrino physics. The forthcoming
results could decisively establish  neutrino oscillations as  the  underlying
physics mechanism for the atmospheric $\numu$ deficit and  provide a 
precise measurement of the corresponding oscillation parameters, $\dmsq23$
and $\sinsq2t23$.

The full potential of the NuMI neutrino beam  can be exploited by 
complementing
the MINOS detector, under construction, with a new detector(s) placed at some
off-axis position and collecting data in parallel with MINOS. The first
phase of the proposed program includes a  new detector,
optimized for  $\nue$ detection, with a fiducial mass of the order of
50 kton and exposed to  neutrino and antineutrino beams. 

An off-axis NuMI neutrino beam offers an unique opportunity to study 
$\numutonue$ oscillations. There will be a very large number of $\numu$'s
oscillating away. Most of the resulting $\nu_{\tau}$'s will be below the 
kinematical threshold for $\tau$ production hence a small  admixture
of  $\nue$'s should be detectable with as small background as 
possible. 

To take full advantage of this opportunity it is necessary to construct a new 
detector capable of the detection and identification of the $\nue$ charged 
current interactions. 
Such a detector must meet several challenges:
\begin{itemize}
\item it must have fine granularity in order to identify the final state 
electrons
\item it must have very large mass to provide maximal 
sensitivity  to the  oscillation amplitude
\item it must have an acceptable cost per unit mass
\item it must be able to operate on surface or under a small overburden, as 
there are no convenient underground locations 
\end{itemize}
The detector should be optimized for the neutrino energy range of $1 - 3~GeV$. 

Identification of the final state electron in a calorimetric detector requires
that the sampling frequency is high, of the order of $1/4-1/3$ of the radiation
length $X_0$. Neutrino detectors must serve as a target and as a detector at
the same time, hence their mass must be maximized. These two requirements lead 
to a conclusion that the absorber should be made out of a low Z material to 
maximize the mass of the detector while maintaining good sampling frequency.
Low Z absorber will lead to a minimal number of the active detector planes, for
a given total mass of a detector, hence it will minimize the cost of the
detector.

A required transverse granularity of the detector is related to the 
local particles density on one hand and to the Moliere radius on the other 
hand. Hadron and electron showers develop over large volumes in a low density
detectors, hence the requirements on the transverse granularity of the detector
will be relatively modest.

The NOvA collaboration\cite{Ayres:2002nm} submitted a proposal to construct a 50~kton sampling detector
built from particle board and liquid scintillator with APD readout. The detector
would be located above ground, with a long baseline of about 800~km and an off-axis
displacement of about 12~km from the main NUMI beamline.

Recently, the collaboration also presented the preliminary design of an attractive
alternative detector based on a totally active liquid scintillator design (TASD).
Simulations of this option showed an improvement in efficiency of almost a factor
two. Like the baseline detector, TASD is a tracking calorimeter with alternating
vertical and horizontal planes of active liquid scintillator contained in PVC extrusions.
Unlike the baseline design, there is no absorber, so TASD is expected to be 85\% active
and 15\% PVC. The overall dimensions are 17.5~m (width) $\times$ 17.5~m (height)
$\times$ 90.4~m (length).

\begin{figure}[tb]
\centerline{\epsfxsize=0.5\textwidth\epsfbox{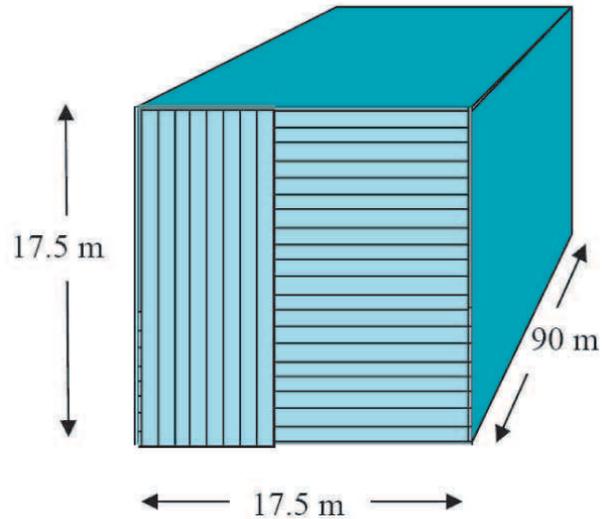}}   
\caption{The proposed FNAL TASD detector composed of 17.5~m long PVC extrusions filled with
liquid scintillator.}
\label{fig:tasd}
\end{figure}

The physics program offered by this design is rather
limited and in direct competition with the approved T2K program.
It is not yet clear how this design can evolve with potential
future neutrino program beyond the NUMI program.

The technology is simple but requires volume instrumentation.
Hence, the scaling properties beyond the currently proposed mass of 25~kton
will be limited by the cost, unless the sampling rate is further reduced.
The sampling rate can be optimized for a given physics application,
as was performed in the NOvA proposal. This necessarily limits
the discovery potential of this experiment beyond what has
been foreseen at the level of the proposal. Given the size
of this project (and the estimated detector cost of about 150~M\$
or about 3 times cheaper than MINOS per kton), the window
opportunity for this technology is rather limited and only
sensible if timely completion with a similar timescale as T2K
can be granted.

This kind of detector is not adequate for a Phase-II program.

\section{MAGNETIZED SAMPLING CALORIMETER}
The MINOS far detector\cite{Gallagher:2004pj} has a total mass of 5.4 kT. It is placed in 
a new cavern dug 713 m underground in the Soudan mine in 
northern Minnesota, about 735 km away from the primary target
at Fermilab. 
The far detector is  made out of two super-modules, each an
8m-diameter octagonal toroid composed of 243 
layers. Each layer is made of a 2.54 cm-thick steel plane and 
1 cm-thick and 4.1 cm-wide scintillator strips grouped in 20- or 
28-strip wide light-tight modules. The scintillator strips 
are made in an industrial extrusion process using 
Styron 663W polystryrene, manufactured by the DOW chemical company,
doped with 1\% PPO and 0.03\% POPOP. A 0.25 mm-thick reflective
layer, made by adding 12.5\% $\rm{TiO}_2$ to polystyrene, is 
co-extruded with the scintillator strips. A 1.4 mm-wide and 2 mm-deep 
groove in the center of the 4.1 cm-side is also made during the 
extrusion process. A 1.2 mm diameter wave length shifting (WLS) fiber
is embedded in the groove during the assembly of the 
scintillator modules. The J-type Y11 multiclad PMMA, non-S WLS fiber 
made by Kuraray and doped with the 
K27 dye at 175 ppm (with maximum intensity emission at 520 nm) is
used. The fibers are optically coupled to the scintillator 
strips with epoxy. The WLS fibers are read out from both ends. 
They are grouped (multiplexed) inside a light-tight box  into 
sets of 8 fibers from strips spaced more than 1 m apart in 
each plane. Each 8-fiber bundle is coupled, using a ``cookie'',
 to a single pixel of 
a 16-pixel  R5900-M16 Hamamatsu photomultiplier (PMT). 
 Thus each PMT reads out 128 fibers; 
one end of each scintillator plane needs 24 pixels. This arrangement
allows us to read out the entire MINOS far detector using only 
1452 PMTs. Since the event rate is small, unambiguous 
event recontruction can be achieved in software despite the 
multiplexing.  An important detector parameter is the 
photo-electron yield for a minimum ionizing particle (MIP) incident at 
right angle to the scintillator strip: the average yield,
measured using a radioactive source
 for each strip during assembly, is about 6 photo-electrons per MIP
 summed from
both sides.  The attenuation over the 8 meter length of the 
strip  is about a factor of 3. 
The detector is magnetized using a coil through a hole in the
center of the planes to an average field of about 1.5 T (2 m away from the
coil). The front end electronics is different for the two detectors
because the event rate in the 8.1 $\mu$-sec neutrino pulse
 in the near detector is far  higher 
than the far detector.  In the far detector the read out electronics
is based on a VA chip from IDE and in the near detector it is 
based on the QIE chip designed at Fermilab.  Simulations show 
that these detectors have a resolution of $60\%/\sqrt{E}$ for 
hadronic showers and $25\%/\sqrt{E}$ for electromagnetic 
showers. Both detectors are being calibrated by cosmic rays 
and a light injection system. A test beam calibration module is 
being used to perform relative calibration between the near 
and far detectors of about 2\% and absolute calibration 
of about 5\%. 

\begin{figure}[tb]
\centerline{\epsfxsize=0.5\textwidth\epsfbox{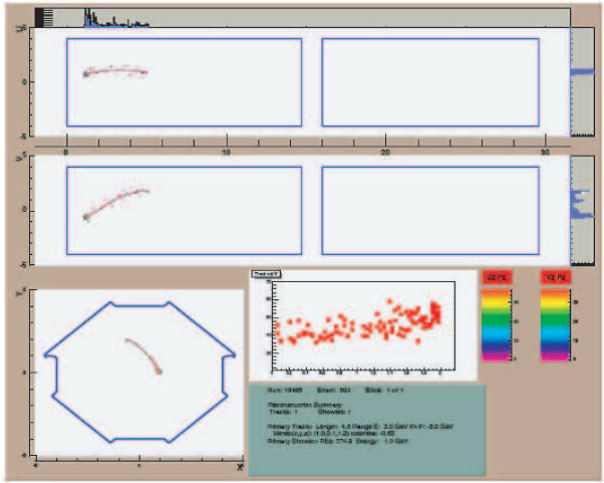}}   
\caption{MINOS: one of the first candidate for contained neutrino interaction.}
\label{fig:minosatm}
\end{figure}

The basic design of this kind of detector represents the 
old iron magnetized technology, with the drawback of volume instrumentation.
Hence, scaling limited by the trade-off between
the number of channels and the sampling rate.

This kind of detector has limited event reconstruction capability, 
and limited electron performance. For background estimation, 
one must heavily rely on Monte-Carlo estimates. It is mostly
adequate for measuring deep-inelastic events and the leading
muon charge. This detector is not adapted
to low energy superbeam or betabeam and cannot address
non-accelerator physics. It is 
most adequate for a neutrino factory, if one is only interested in wrong sign muons\cite{DeRujula:1998hd}.

\section{EMULSION-HYBRID}
The basic performance of an OPERA-like
detector at the Neutrino Factory in reconstructing neutrino
interactions was addressed in Ref.\cite{Autiero:2003fu}.

The experiment uses nuclear emulsions as high
resolution tracking devices for the direct detection of the $\tau$
produced in the $\nu_{\tau}$ CC interactions with the target.

\begin{figure}[h!]
\begin{center}
\epsfig{file=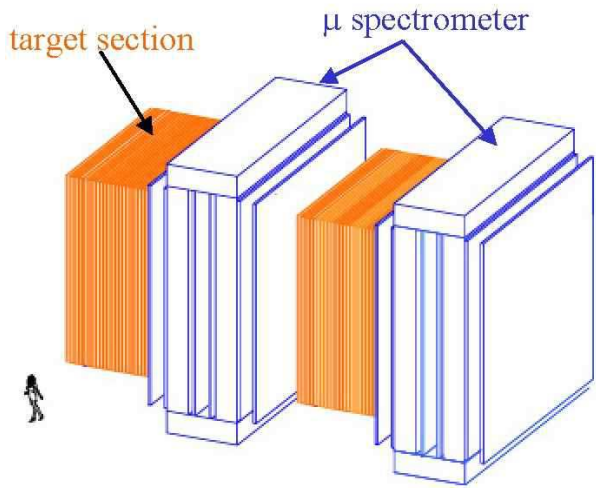,width=0.5\textwidth}
\end{center}
\vspace{-0.5cm}
\caption{\label{fig:opera_cngs}  \it OPERA at CNGS.}
\end{figure}

OPERA is designed starting from the Emulsion Cloud Chamber concept
which combines the high
precision tracking capabilities of nuclear emulsions and the large
mass achievable by employing metal plates as a target. The basic
element of the ECC is the cell which is made of a 1~mm thick lead
plate followed by a thin emulsion film.  The film is made up of a pair
of emulsion layers $50~\mu$m thick on either side of a $200~\mu$m
plastic base.  Charged particles give two track segments in each
emulsion film.  The number of grain hits in about $50~\mu$m
($15$-$20$) ensures redundancy in the measurement of particle
trajectories. By piling-up a series of cells in a sandwich-like
structure bricks can be built, which constitute the detector element
for the assembly of massive planar structures (walls). A wall and
its related electronic tracker planes constitute a module. A
supermodule is made of a target section, which is a sequence of
modules, and of a downstream muon spectrometer. The detector consists
of a sequence of supermodules (see Fig.~\ref{fig:opera_cngs}).

The signal of the occurrence of a $\nu_\tau$CC interaction in the
detector target is identified by the detection of the $\tau$ lepton in
the final state through the direct observation of its decay
topology. A $\tau$ may decay either into the lead plate where the
interaction occurred (short decay) or further downstream (long
decay). For long decays, the $\tau$ is detected by measuring the angle
between the charged decay daughter and the parent direction. The
directions of the tracks before and after the kink are reconstructed
in space by means of the pair of emulsion films sandwiching the lead
plate where the interaction occurred. A fraction of the short decays
is detectable by measuring a significant impact parameter (IP) of the
daughter track with respect to the tracks originating from the primary
vertex.

The detection of the $\tau$ decay and the background reduction benefit
from the dense brick structure given by the ECC, which allows the
electron identification through its showering, the pion and muon
separation by the $dE/dx$ measurement method, and the determination of
the momentum of each charged particle employing techniques based on
the Multiple Coulomb Scattering. All these methods are discussed in
the following.

Electronic detectors placed downstream of each emulsion brick wall are
used to select the brick where the interaction took place (to be removed for 
the analysis) and to guide the emulsion scanning. The target
electronic detectors are also used to sample the energy of hadronic
showers and to reconstruct and identify penetrating tracks.

Overall, this detector is a hybrid between a sampling calorimeter and emulsions.
Future application of this technology will be limited by its handling complexity
and by the scanning load which is perceived as the bottle neck. 
The mass scaling is limited by cost of emulsions and
event statistics is limited by emulsion scanning.
It has an excellent long-lived track finding capability which is
adequate for tau and charm identification. A possible
physics program at a neutrino factory has been developed in Ref.\cite{Autiero:2003fu}.


%

\section{CONCLUSION}

From the considerations developed in this paper, we draw the following conclusions:
\begin{enumerate}
\item Not all technologies can satisfy the requirement
of the general purpose and versatile detector, capable of simultaneously addressing non-accelerator
and accelerator physics programs. 
\item Only water Cerenkov detector and liquid Argon TPCs
address in a satisfactory (and in fact in a complementary way, see e.g. Ref.\cite{Rubbia:2004yq}) solar, supernova, atmospheric
neutrino and nucleon decay searches. 
\item Superbeams and betabeams of subGeV and GeV energies can be coupled
to water Cerenkov detector and liquid Argon TPCs. However, the high-$\gamma$ betabeam
with energies above the GeV can only be efficiently coupled to a liquid Argon TPC, 
because Water Cerenkov are optimal for single ring events.
\item The neutrino factory necessarily requires magnetized detectors to measure
the charge of the leptons in order to discriminate
between signal and background. A magnetized liquid Argon TPC or a non-magnetized
liquid Argon TPC with an external muon spectrometer can be considered,
aiming at identifying  and measuring all three charged lepton flavors
produced in charged current interactions and at measuring 
their charges to discriminate the incoming neutrino helicity.
\item Magnetized sampling calorimeters can be used at a neutrino factory for muon detection.
Emulsion hybrid detectors can be in principle
considered for tau detection, although serious handling issues may arise.
These detectors cannot address the non-accelerator physics program.
\end{enumerate}

The megaton Water Cerenkov detector may represent a conservative approach given
the experimental knowledge gathered so far. However,
the ``100 kton liquid Argon'' is certainly a challenging but potentially very performing design.
Given the foreseeable timescale of more than a decade for the 
next generation of massive underground detectors, it is our conviction that
a certain level of challenge represents the most attractive
way towards potential progress in the field.

\section*{Acknowledgments}
We thank F.~Cervelli for organizing such a wonderful workshop in Elba.
We acknowledge A.~Ereditato for useful discussions.

\end{document}